\documentclass{article}

\begin{document}

\title{Kinetics of nucleation under
the break of the blow-up growth}
\author{Victor Kurasov}
\date{Victor.Kurasov@pobox.spbu.ru}

\begin{abstract}
Kinetics of the first order phase
transition has been
investigated.
The case when the droplets
sizes have a limit value is considered.
An analytical theory for such  process has
been constructed and all main
characteristics  of the process
have been determined by rather
simple approximate analytical formulas.
The limit for a droplets size
violates the avalanche consumption of the mother
phase
which is the basic feature of
the first order phase transition and
have been actively used in previous
publications. Now
one can not state that the main
quantity of the already
condensated substance is
contained in the droplets formed under
conditions unperturbed by nucleation process.
This requires to
introduce some new methods of solutions for
kinetic equation which
are given in the present paper.

\end{abstract}

\maketitle

\section{Introduction}

 The most
interesting feature
of the first order phase transition
is the global kinetics of
transformation of a mother phase
into a state of objects of a new phase.
Traditionally
the theoretical description of the
first order phase transition
kinetics is given on example of nucleation
\cite{Zeld}.
Then the mother phase is a
supersaturated vapor and
the new phase is a liquid phase.
The driving force  of the first order phase transition
is initiated by
metastability which is
traditionally described by the value of a
supersaturation $\zeta$.
The last is the ratio of the density of
surplus vapor and the density of
the saturated vapor minus one unit.

Kinetics of nucleation has been already
investigated in the case
of the ordinary (unlimited)
phase transition. In that case the
property of the blow-up growth has
been actively used \cite{book1}.  This
property means that the rate of
growth
of the molecules number  inside the droplet
for a given droplet rapidly
increases in time. It grows so rapidly,
that even acceleration of
the rate of growth also grows in time.
This property leads to two
important consequences, which
allow to give an analytical
description of the nucleation period.
These consequences are the
following:
\begin{itemize}
\item
The nucleation period (i.e.
the period of  rather
intensive appearance of supercritical
droplets) is well localized in time.
\item
The droplets appeared in the first
moments of nucleation period
contain the main quantity of
molecules (in comparison with other
droplets) during the nucleation period.
These first moments can be
called as the pre-nucleation
period. During the pre-nucleation
period the droplets don't
essentially perturb
the state of the system.  This
allows to calculate the
characteristics of droplets appeared
during the pre-nucleation period.
\end{itemize}

For example, the regime of the
unlimited
free molecular growth has the
mentioned feature. Namely here the
number of molecules  $\nu$ inside the
supercritical embryo (i.e. inside
the droplet) grows in time $t$
as
\begin{equation} \label{1}
d(\nu)^{1/\alpha}/ dt
=
\zeta/ \tau, \ \  \  \alpha=3,
\end{equation}
where $\tau$ is
a characteristic constant which has a
sense of a mean
time between collisions of a
given molecule in saturated vapor
with other molecules.
This type of growth is wide spread in nature.
Later we shall
use this type of growth as a
model for concrete formulas.

In previous theories for nucleation
kinetics (see rewiev in \cite{book1})
it was supposed that the
sizes of droplets can become in
principle infinitely big.
Here we consider the opposite situation.
In many systems the process of
the droplets growth
is limited by some artificial
restrictions. The number $\nu$ of
molecules inside the droplet can
not be greater than some value
$\nu_{lim}$, which is an external
parameter of the theory. We imply that
$\nu_{lim}$ strongly exceeds the
number $\nu_c$ in the critical embryo
under the supersaturation which
corresponds to relatively intensive
appearance of
droplets (this condition will be
specified).    The
existence of a limit  size takes
place in the processes of cementation, of
condensation in an external  lattice,
morphological transitions,
etc. This type of transition is rather
wide spread in nature.
Kinetics of this process will be
investigated in this paper.

The limit of droplets sizes leads
to the loss of the blow up property of
growth in the global kinetics of
nucleation. From the first point
of view it seems that the
introduction of a limit size can not
essentially change the approaches
of description, but it is an
illusion and below some principally
new methods of solution of kinetic
equations have to  be presented.

This publication continues
the set of articles devoted to the
construction of the global kinetics
of a phase transition (see
\cite{book1}).
To grasp the current approach it
would be useful to come
through these papers.
We use all physical assumptions
from these papers except
explicitly mentioned ones.
All definitions of physical
characteristics can be taken also from
\cite{book1}.

Traditionally two characteristic types
of external conditions are
considered. The first type is a
decay of a metastable state
\cite{Kuni-Grinin}. Here at the
initial moment of time we have the
metastable phase with no
objects of a new phase in the system.
Later there will be no external
influence on the system.
The
process of nucleation begins
and the appearance and growth of
droplets cause the fall of
supersaturation which stops the further
intensive nucleation.

The second type is nucleation under
the smooth variation of
external conditions. At first
moments of time the increase  of
supersaturation by the action of
external conditions is greater
than the decrease due to the mother
phase consumption by the
droplets. But the vapor consumption
has a very rapidly increasing
intensity and later the mentioned
decrease will overtake the
mentioned increase and the
supersaturation will go down. The
process of intensive nucleation will stop.
We are going to describe kinetics of
the process in both situations.

We shall use the following law of
droplets growth for $\rho \equiv
\nu^{1/3}$:
$$
\frac{d\rho}{dt}
= \zeta/\tau \ \ \ for \ \
\rho<\rho_{lim} \equiv \nu_{lim}^{1/3}
$$
$$
\frac{d\rho}{dt} = 0
\ \ \ for \ \
\rho \geq\rho_{lim} \equiv \nu_{lim}^{1/3}
$$

Rigorously  speaking it would necessary to
determine the variable
which has the velocity of growth independent
on the value of this
variable. We shall act in another manner:
at first we suppose that
there is no limitation on growth and then
calculate the number of
molecules  in a droplet by a "curved" formula.

Here it is possible to introduce the
following variables: The
variable $y$ is the number of molecules in
the droplet which is
growing without any limit by an initial
law (\ref{1}). The variable
$z$ is the value of $y$ for droplet which
was born at "initial" moment  of
time. Certainly, for every droplet
$\nu<z$. The variable $x$ is
defined as $y=z-x$. Then we see the
correspondence between $t$ and
$z$ and the correspondence
between $t$ and $x$. The values $z$ and
$x$ are now equivalent.

The number of molecules $\nu$ inside
the droplet has to be
calculated now as
$$\nu = (z-x)^3 \ \ \ for \ \ z-x<\rho_{lim}$$
and
$$\nu = \nu_{lim} \ \ \ for \ \ z-x>\rho_{lim}$$
We see that here there is no unique
analytic dependence and this will seriously
complicate the evolution equation.

We have also to mention that the property of
the effective size of
growth which was observed in recent
publication \cite{smooth} in
construction of kinetics for the
law (\ref{1}) with small $\alpha
\leq 1$ also doesn't take place.
For droplets with\footnote{Here
$\nu$ is imaginary value calculated purely according
to (\ref{1}).} $\nu \gg
\nu_{lim}$ such a property is observed,
but for droplets with $\nu
< \nu_{lim}$ is the opposite property of
the blow-up regime takes
place. Namely the situation
where both regions are important in
substance balance is interesting.

All analytical constructions are
based on the following approximation for
the  rate of nucleation $I$ as
a function of
 supersaturation:
\begin{equation}\label{+++}
I(\zeta) = I(\zeta_0) \exp(-p (\zeta - \zeta_0))
\end{equation}
Here $\zeta_0$ is the base of
decomposition and the
positive parameter $p$ is
defined as
$$
p = - \frac{dF_c}{d\zeta}|_{\zeta= \zeta_0}
$$
where $F_c$ is the free energy of a
critical embryo.
The last approximation is not more
than the Klapeiron-Klausius
approximation, but it is written
for special variables. The
validity of this approximation
has been discussed in
 \cite{book1}.

\section{Decay of metastable state}

Now we shall consider the
situation of decay. The initial value of
supersaturation will be marked
by $\Phi$. The base of
decomposition  $\zeta_0$ for function $\zeta$ can
be chosen as $\Phi$.

We skip the derivation of kinetic
equation because it is quite
similar to the ordinary case
without the limit size and can be
seen in \cite{PhysRevE94} or in \cite{PhysicaA94}.
The kinetic equation for the
evolution can be written in the
following form:
\begin{eqnarray}
\Phi = \zeta (z) +
\Theta(z - z_{lim})\{
A \int_{z-z_{lim}}^z (z-x)^3
\exp(p(\zeta (x) - \Phi)) dx +
\nonumber \\  \label{kin}
\\ \nonumber
A \nu_{lim} \int_0^{z-z_{lim}}
\exp(p(\zeta (x)- \Phi)) dx
\}+
\Theta(z_{lim}-z)
A \int_0^z (z-x)^3
\exp(p(\zeta (x) - \Phi)) dx
\end{eqnarray}
Here $A$ is the amplitude value of
spectrum, $\Theta$ is the
Heavisaid's function, $z_{lim} =
\rho_{lim}$. Solution of this
equation will give the
behavior $\zeta$ as a function
of $x$. It allows to calculate
the number of droplets
(in renormalized units)
$$
N(z) = \int_0^z \exp(p(\zeta(x) - \Phi)) dx
$$
appeared until the moment
$t(z)$ (i.e. until the moment when the
imaginary front side of the spectrum
attains the value $z$). The most
interesting value is the
total number of droplets $N(\infty)$ and
the error in this value
will be the measure of the accuracy of
approximate description.

In contrast to the  decay with the
unlimited  growth  we shall
have here the several stages of the
process of decay and the long
continuous tail. The meaning of
these parts of the process
will be clear later
and now we shall start with
investigation of the first period and have
to put initial conditions. Namely,
$t=0$, $\zeta = \Phi$ and there
is no droplets in the system.

Now we shall introduce the
value of natural length $\Delta_1 x$ as
the characteristic width of
spectrum without the limit of growth.
For this situation we have the
following kinetic equation
$$
\Phi = \zeta (z) +
A \int_0^z (z-x)^3  \exp(p(\zeta (x) - \Phi)) dx
$$
which results in the following
characteristic width
of the size spectrum
$$
f \sim
\exp(p(\zeta (x) - \Phi))
$$
namely
\begin{equation}\label{Delta}
\Delta_1 z = (\frac{4}{pA})^{1/4}
\end{equation}

One can see that $\Delta_1 z$ is a
function of $\Phi$ since
$p=p(\Phi)$, $A=A(\Phi)$.
It is possible to see three
situations:

(A) $\Delta_1 z  \ll z_{lim}$,

(B) $\Delta_1 z  \gg z_{lim}$,

(C) $\Delta_1 z  \approx z_{lim}$.

\subsection{ The case $\Delta_1 z  \ll z_{lim}$}

In this case kinetic equation
(\ref{kin}) can be reduced to
$$
G(z) =
A \int_0^z (z-x)^3  \exp( - p G(x)) dx
$$
for the function
$$
G = A
\int_0^z (z-x)^3  \exp(  p (\zeta - \Phi) ) dx
$$
which has the sense of the number of
molecules inside the new phase.

This equation is known from
unlimited case and can be solved by
iterations:
$$
G_0 = 0
$$
$$
G_{(i+1)}(z) =
A \int_0^z (z-x)^3  \exp( - p G_{(i)}(x)) dx
$$
Calculation gives
$G_{(1)} = Az^4/4$ and already
the second iteration gives the
practically
accurate number of droplets
$$
N_{(2)} (\infty) \equiv
\int_0^{\infty} \exp(-pG_1(x))
dx= (\frac{4}{pA})^{1/4} B,
\ \ \ B
\equiv \int_0^{\infty} \exp(-x^4) dx = 0.9
$$
The relative error
$$
\epsilon (\infty) =
\frac{|N(\infty) -N_{(2)} (\infty) | }{N(\infty) }
$$
is less than $0.05$ in every case
which can be analytically
proven.

The  high accuracy is also attained also for
$$
L_i =
\frac{3!}{i!(3-i)!} A
\int_0^{\infty} x^i \exp(-p G) dx
$$
calculated on the base of the first iteration i.e
for
$$
L_i \approx L_{i\ (2)} =
\frac{3!}{i!(3-i)!} A
\int_0^{\infty} x^i \exp(-p G_1) dx
$$
These values  will be important later.

But this doesn't complete the investigation
as in the case of
unlimited growth. Earlier or later the
droplets sizes attain $\nu_{lim}$
and the droplets stop to grow. Then the final
value of the supersaturation will
be
$$
\zeta_{fin\ 1} = N(\infty) \nu_{lim}
$$
Here index $1$ points that we investigate
the first part of the
nucleation period.

The time of attaining $\zeta_{fin}$
can be easily calculated on
the base of monodisperse
approximation\footnote{The
validity of monodisperce approximation
can be proved analytically.}. The last approximation
leads to  equation
\begin{equation} \label{mono}
A N(\infty) z^3 = \Phi - \tau \frac{dz}{dt}
\end{equation}
which can be easily integrated
(as the first order differential
equation without explicit dependence on the
argument):
$$
\int_0^z \frac{dx}{\Phi - A N(\infty) x^3} = t
$$
The integral can be taken analytically
and gives $z(t)$. Condition
$$
 z^3 (t_{fin}) =  \nu_{lim}
 $$
 gives the value of the
 end $t_{fin}$ of the first part of
 nucleation. One
 can compare
$t_{fin}$  with the duration
of intensive nucleation
during the first part, i.e. with
$$
\Delta_1 t = \Delta_1 z \tau  /
\Phi
$$

When the
inequality $\Delta_1 z \ll z_{fin}$
isn't too strong
then the equation (\ref{mono})
isn't suitable because the monodisperse
approximation lying in the
base of (\ref{mono}) isn't valid even
at $z \approx z_{fin}$. Then
one has to use instead of (\ref{mono})
the following equation
\begin{equation} \label{nonmono}
\sum_{i=0}^3 L_i  z^3 = \Phi - \tau \frac{dz}{dt}
, \ \ \  z|_{t=0} = 0
\end{equation}

The methods of solution  of  this equation are
quite similar to the methods of
solution of (\ref{mono})
(they are presented in \cite{book1}).

While approaching to $z_{fin}$ one has
to take into account that
$L_i$ begin to change
because integration can not
be done out of $z<z_{fin}$.
These changes can be taken into
account by the perturbation
technique (because the main influence is
given by $L_0$ which remains unperturbed).

Several important statements can
be proved here. They are:
\begin{itemize}

\item
(A) The process of nucleation takes place in the
quasistationary conditions. The time $t_1$ of
essential perturbation of the droplets
distribution, which can be estimated as $\Delta
z_1 \tau / \Phi $ strongly exceeds the time $t_s$
of establishing the stationary state in the near
critical
region (more accurate in the region\footnote{
To calculate $t_s$ see \cite{Zeld}, \cite{grinints}.
} $\nu
< (2 \div 3 ) \nu_c$).

\item
(B) The main role in the vapor consumption is
played by the strongly supercritical droplets,
i.e. by the droplets
with $\nu > (2 \div 3) \nu_c$. These
droplets grow regularly with the mentioned law of
growth.

\item
(C) The process of intensive nucleation takes
place at $\Phi \geq \zeta \geq
\Phi - 1/p $.

\item
(D) Later the new periods of nucleation (may be
not so intensive but long) will take place.

\end{itemize}

The statements (A), (B) are necessary for
construction of the kinetic equation. The
statement (C) allows to use the approximation
(\ref{+++}). The statement (D) requires to
continue the consideration of the nucleation
process which will be done below.

The last step to do is to give approximation for
the number of droplets. After the
evident rescaling the
kinetic equation can be presented as
$$
G = 4 \int_0^z (z-x)^3 \exp(- G(x)) dx
$$
The number of droplets appeared until $z$ can be
  calculated as
$$
N(z) = \int_0^z \exp(-x^4) dx
$$
We shall use the following approximation
$$
N_{appr} (z) = \Theta(0.91-z) z + \Theta(z-0.91)
0.91
$$
The relative error $|N_{appr} - N|/ N$
as function of $x$ is drawn in Figure 1. We see
that it is small.

\subsection{ The case $\Delta_1 z  \gg z_{lim}$}

This case is much more simple than the previous
one. Here the kinetic equation can be written as
$$
\Phi = \zeta (z) +
A \nu_{lim} \int_0^{z-z_{lim}} \exp(p(\zeta (x)-
\Phi)) dx
$$
or
$$
G = A \nu_{lim} \int_0^{z-z_{lim}} \exp( - p G(x))
dx
$$
for
$$
G = A \nu_{lim} \int_0^{z-z_{lim}} \exp(p(\zeta
(x)- \Phi)) dx
$$
It can be solved very easy. Namely, after the
differentiation we come to
$$
\frac{dG}{dx} = A \nu_{lim}  \exp( - p G(z))
$$
which  can be easily integrated
$$
\frac{1}{p}( \exp(p G(z)) - 1)  = A \nu_{lim} z
$$
The spectrum has the form
$$
f \sim A \exp(-p G(z))
$$

One can see that here the number of
total droplets
$$
N(\infty) = \int_0^{\infty}  f (x) dx
$$
is infinite. It is certainly a wrong result. The
reason is inapplicability of approximation
(\ref{+++}). The head of the spectrum is
described quite satisfactory, but the tail isn't
described well. Really at $n=n_{\infty}$ the
nucleation according this approximation doesn't
stop. To describe the tail we have to use the
precise equation for the stationary nucleation
rate
$$
I_s  = Z \exp(-F_c)
$$
where $Z$ is the Zeldovitch factor \cite{Zeld},
$F_c$ is the free energy of the critical embryo,
taken in some approach but without approximate
decomposition (\ref{+++}). For example, in
capillary approximation
$$
F_c = \frac{1}{3} a \nu_c^{2/3}, \  \
\nu_c^{1/3} = \frac{2 a }{3 \ln(\zeta+1)}
$$
where $a$ is renormalized surface tension. Then
$$
\int_0^t  \nu_{lim} I_s (\zeta(t')) dt'
=
\Phi - \zeta(t)
$$
or
$$
 \nu_{lim} I_s (\zeta(t))
=
 - \frac{d \zeta(t) }{dt}
$$
The last equation can be easily integrated
$$
\int_{\Phi}^{\zeta}
\frac{d\zeta}{ \nu_{lim} I_s (\zeta(t))  }  =
t
$$
or
$$
\int_{\Phi}^{\zeta}
\frac{ \exp (\frac{4 a^3}{27 \ln^2(\zeta+1)})
d\zeta}{ \nu_{lim} Z(\zeta) }    =
t
$$

One can get an analytical solution for the
tail  of spectrum having noticed that at small
$\zeta$
one can decompose $\ln(\zeta +1) \approx \zeta$.
 The values of $\zeta$ can not be too
small, at least $\nu_c(\zeta) < \nu_{lim}$ and
the behavior of $Z$ isn't singular. So we can
reduce the last equation to
$$
B_2 \int
 \exp (B_3  \zeta^{-2})
d\zeta    =
t +B_1
$$
with  known constants  $B_i, i=1,2,3$.
The integral can be reduced to the error function.
This solves the problem of explicit solution in
terms of standard
special functions.

Here one has no need to observe the property (B)
because this property begin to be violated only
when $\nu_{lim}^{1/3}$ strongly exceeds $\Delta
z_1$ and then practically all droplets attain the
final values. As for the quasistationarity
(property
(A)) we see that $|d \zeta / dt| $ decreases in
time and if (A) is observed in the initial moment
of time it will be observed also later. The
absence of the property (C) was taken into
account just above. Instead of (D) one can state
that this stage will be the last stage of
the whole nucleation
process in the system.

\subsection{ The case
$\Delta_1 z  \approx z_{lim}$}

The intermediate case is the most complex one
and has to be investigated on the base of the already
presented solutions. Let the kinetic equation be
rescaled to have
$$
G =
$$
$$
\Theta(z - z_{lim})\{
4 \int_{z-z_{lim}}^z (z-x)^3
\exp(- G(x)) dx +
4 \nu_{lim} \int_0^{z-z_{lim}}
\exp( - G(x)) dx
\}+
$$
$$
\Theta(z_{lim}-z)
4 \int_0^z (z-x)^3  \exp( - G(x)) dx
$$
Certainly, here $z_{lim}$ will be also rescaled.

Now we shall give the approximate method to solve
this equation. The value $G$ can be presented as
$$
G =g_+ + g_-
$$
where
$g_+$ is the rescaled number of molecules in the
droplets which have already attained $z_{lim}$,
$g_-$ is the rescaled number of molecules in all
other droplets.

For further
constructions we have to write approximations for
$g_-$ and $g_+$. We shall start with $g_-$

The function $g_-$ can be easily found on the
base of iteration procedure presented for the
case $\delta_1 z  \ll z_{lim}$. Then
$$
g_- = \Theta(z-\Delta_1 z)
4 A \int_0^{\infty} (z-x)^3 \exp(-g_- - g_+) dx +
\Theta(\Delta_1 z - z )
4 A \int_0^z (z-x)^3 \exp(-g_- - g_+) dx
$$
Here $A$ is the amplitude which can  differ from
$1$ due to the possible big value of $g_+$.
At the current moment
this remark isn't too clear, but in any case we can
say that for further
purposes it is convenient to conserve $A$
here.

Now we shall explain the last relation.

At $z<z_{lim}$  we have $g_+ = 0$ and
$$
g_-(z) =  4 A \int_0^z (z-x)^3 exp(-g_-(x)) dx
$$
It is important  that
the last equation  is the closed one
and it is identical to
the case of unlimited  growth.
So, it can be successfully solved
by iterations. Particularly,
$g_{-(0)} = 0 , g_{-(1)} = A z^4 $, etc.

At $z\geq z_{lim}$ we have
$$
g_- = 4 A
\int_{z-z_{lim}}^{z} (z-x)^3 \exp(-g_-(x) - g_+(x))
dx
$$

Now we give the qualitative picture of phenomena.
At the first moments of time the spectrum in the
region $\rho <\rho_{lim}$ is being formed and
$g_+=0$ but $g_-$ is growing rather rapidly $\sim
z^4$. Later the substance is going to be
accumulated in droplets with $\rho = \rho_{lim}$.
The growth of $g_+$ occurs.

If the size of cut-off $z_{lim}$ is essentially
smaller than $\Delta_1 z \equiv A^{-1/4}$ then
$g_-$ accumulates not so many molecules of
substance
$g_- = g_{- (in)} \equiv Az_{lim}^4 \ll
1$ and later due to the decrease of nucleation
rate the value of $g_-$ will also decrease, i.e.
$g_- < g_{-(in)}$ In this case $g_+$ grows as
$$
G_+ \approx z_{lim} 4 A \int_0^{z-z_{lim}}
\exp(-g_+(x)-g_-(x)) dx
$$
It will grow slower  than
$$
g_{+(in)} = z_{lim} 4 A (z-z_{lim})
$$
This
allows to state that
$$
\frac{d g_+}{dz} <  4 A z_{lim}^3
$$
Then the
variation $\delta g_+$ in $g_+$ during "the time" of
establishing of the quasistationary
state in the region
$[0,z_{lim}]$ can be estimated as
$$\delta g_+ < \frac{dg_+}{dz} z_{lim} =
4 A z_{lim}^4 \ll 1$$
In the expression for $g_-$ for $z>z_{lim}$
the last inequality allows to
take $\exp(-g_+)$ away from the
integral.

Then approximately
$$
g_-(z) = 4 A
\exp(-g_+(z))
\int_{z-z_{lim}}^{z} (z-x)^3 \exp(-g_-)
dx
$$
and on the base of initial iteration
approximation
$$
g_-(z) =  A
\exp(-g_+(z))
z_{lim}^4
$$

Now we shall get expression for $g_+$.
We shall get for $g_+$ the following approximate
equation
$$
g_+ = 4 A z_{lim}^3 \int_0^{z-z_{lim}}
\exp(-g_-(x) - g_+(x))
dx
$$
Here
$z$ is the current
"moment of time", $z_{lim}$ is the
"moment" when the spectrum attains $z_{lim}$.

Here we are interested in the
other characteristic scales of time
which are greater than $1/(4Az_{lim})$.
At these times we can
suppose that $\exp(-g_-)$ is
approximately constant and take it
out of the integral.
Then approximately
$$
g_+ = 4 A z_{lim}^3 \exp(-g_-) \int_{0}^{z-z_{lim}}
\exp( - g_+(x))
dx
$$
When $g_-$ is essential
(i.e. $\exp(-g_-)$ is small) then $g_+$
isn't necessary and we can
take out $\exp(-g_-)$. But when $g_-$
is small, then $\exp(-g_-) = 1$
and we  can  also take it out.
The substance consumption
in $g_-$ occurs in the avalanche manner.

We can get solution of this equation quite
analogously to the case $\Delta_1 z \gg z_{lim}$
but here the initial  conditions will be another
$$
g_- |_{x=z-z_{lim}} = 0
$$
Solution will be
$$
\frac{ d g_+}{dz}
= 4 A z_{lim}^3 \exp(-g_-)
\exp( - g_+)
$$
and
$$
\exp(g_+) - 1 = 4 A z_{lim}^3
\exp(-g_-)(z-z_{lim})
$$
The spectrum will be
$$
f = 4 A \exp(-g_-)/\exp(g_+)
$$
or
$$
f = 4 A \exp(-g_-)/
( 1 + 4 A z_{lim}^3
\exp(-g_-)(z-z_{lim}))
$$

In the initial approximation we consider
$A=1$ and get
$$
f = f_0 \equiv 4 \exp(-z_{lim}^4 )/
( 1 + 4 z_{lim}^3
\exp(-g_-)(z-z_{lim}))
$$
Then we take $g_- = z_{lim}^4$ and get
$$
f = f_0 \equiv 4 \exp(-z_{lim}^4 )/
( 1 + 4  z_{lim}^3
\exp(-z_{lim}^4)(z-z_{lim}))
$$

It is necessary refine expression
for the spectrum obtained in
the zero approximation.
We have to take into account that $g_+$  is
formed under the influence of $g_-$.
But the spectrum  in the last
expression is written through $g_-$ (but not
through $g_+$). Here
is the difficulty. We shall
come it over via the special
approximate approach.
At first we assume that $g_-$ (but not the
$g_+$) is formed with  the
renormalized intensity. We take
$\exp(-\alpha z_{lim}^4)$
instead of $\exp(-z_{lim^4})$. It is
evident that we do this only in the
first factor which is
responsible for the influence of $g_-$.
So, we get
$$
f =  4 \exp(-\alpha z_{lim}^4 )/
( 1 + 4  z_{lim}^3
\exp(-z_{lim}^4)(z-z_{lim}))
$$
For $\alpha$ we have to take the
renormalized amplitude. The
natural candidate for this amplitude is
$f_0$. As the result we
come to  the following expression for
spectrum
$$
f_1 = F(f_0) \equiv f_0 K
$$
where
$$
K =
\frac{\exp(-f_0 z_{lim}^4)}{\exp(-z_{lim}^4)}
$$
$$
f_0 \frac{\exp(-z_{lim}^4)}{1+4 z^3_{lim}
\exp(-z_{lim}^4)
(z-z_{lim})}
$$

Now let us realize what have we
done. In reality we had to take
into account the inverse  influence,
i.e. to take into account
that  $g_+$ is formed under the influence
of $g_-$, but not the
influence of $g_+$ on $g_-$.
To take into account the inverse
influence we shall use the inverse
transformation $F^{-1}$. So,
approximately
$$
f = F^{-1}(f_0)
$$
We can calculate this transformation rather easy
$$
f = F^{-1} (f_0) = \frac{f_0}{K}
$$

This will solve the problem of the adequate
construction of the approximate spectrum.

The error of this approximation
in the total droplets number is drawn in
Figure 2. Certainly, we can not calculate here
the total number of droplets (it will be infinite
and has to be corrected by the way
described in the previous
subsection). As the final value $z_{fin}$ the
magnitude $10$ has been chosen.
The last belongs to the
asymptotic  period. One can see that the relative
error in the droplets number is small.

Statements (A), (B), (C) take place here.
Statement (D) isn't necessary.

\subsection{Further nucleation}

The case $\Delta_1  z \ll z_{lim}$ requires the
analysis of the further nucleation.
This analysis isn't too complex. On the base of
$\zeta_{fin\ 1}$ we can calculate the length
$\Delta_2 z$ as
$$
\Delta_2 z = \Delta_1 z (\zeta_{fin \ 1})
$$
i.e. we use the same formula but with $\zeta_{1\
fin}$ instead of $\Phi$.
In expression (\ref{Delta})
there is an amplitude of spectrum $A$
which is a very sharp function of
supersaturation. So,
we see that
$$
\Delta_1 z \ll \Delta_2 z
$$
This strong inequality allows to neglect the
time interval $[0,\Delta_1 z \tau/\Phi]$
in the nucleation at $\zeta_2$. Then the
situation is reduced to the nucleation with
initial supersaturation $\zeta_2$. We have to
repeat the steps of the three previous sections.
Again we  can come to the
possibilities (B) and (C).
The possibility (A)
cannot take place because the inequality
$$
\exp(N_1 x_{lim}^3 /4 )  \ll x_{lim}
$$
doesn't take place. Here the l.h.s. is $\Delta_2
z$ in rescaled units, $N_1$ has to be put to
$0.9$, the coefficient
$4$ in denominator came from the power
$1/4$ in (\ref{Delta}), the parameter
$x_{lim}$ has to satisfy
$x_{lim} > 2 \div 3$.

So, we see that the structure of spectrum isn't
too
complex - the plateau, the region
of partial collapse of surplus substance
 and may be the tail.

The analysis of nucleation
after instantaneous creation of a
metastable state has been completed.

\section{Nucleation under the
smooth behavior of external
conditions}

The straightforward generalization of kinetic
equation to the case of smooth change of external
conditions leads to the following kinetic
equation
\begin{equation}\label{k1}
\Phi(z) = \zeta (z) +
\end{equation}
$$
A \int_{z-z_{lim}}^z (z-x)^3
\exp(p(\zeta (x) - \Phi_*)) dx +
A \nu_{lim} \int_{-\infty}^{z-z_{lim}}
\exp(p(\zeta (x)- \Phi_* )) dx
$$
Here $\Phi(z)$ is ideal supersaturation, i.e. the
supersaturation which should take place
in the system without
any processes of vapor consumption and heat
release effects. The value of $\Phi$ at some
characteristic moment $t_*$ will be marked as
$\Phi_*$. We shall choose  $t_*$ later.
The number of droplets
(in renormalized units) will be calculated as
$$
N(z) = \int_{-\infty}^z
\exp(p(\zeta(x) - \Phi)) dx
$$

Under the unlimited growth of droplets the period
of nucleation is well localized in time. Moreover
it is rather short in time which allows to
linearize the ideal supersaturation  during this
period.
Here the process of nucleation cannot be
localized in time, it is seen simply by the fact
that the number of droplets in a liquid phase
is
limited from above by
$N \nu_{lim}$.
Then to compensate the "action" of external
conditions it is necessary to have nucleation
again and again.

To present concrete calculations we
suppose that the ideal supersaturation can be
linearized
$$
\Phi(z) = \Phi_* + \frac{d\Phi(z)}{dz} |_{z=0} z
$$
but now is is no more than  a model. Later we
shall see how to construct the theory for
rather arbitrary behavior $\Phi(z)$.

But even with this linearization
the situation here can not be solved by a simple
generalization of the ordinary iteration procedure
because here process isn't limited in time.

\subsection{Rescaling}

Kinetic equation (\ref{k1})
can be written in the following form
$$
g= \int_{-\infty}^{z-x_{lim}} x_{lim}^3 f(x) dx
+
\int_{z-x_{lim}}^z (z-x)^3  f(x) dx
$$
with $x_{lim} = z_{lim}$ and a spectrum $f(x)$
$$
f(x) = a \exp(bx - g(x))
$$

It is essential that
after the evident rescaling
one can put $a=1$ and $b=1$.
The third parameter $x_{lim}$ remains and this
leads to the absence of universality.

Restriction $a=1$, $b=1$ doesn't lead to the
special meaning of the point $z=0$ (in
homogeneous condensation $z=0$ was the point of
the maximum of supersaturation, in heterogeneous
condensation $z=0$ was the point where the half
of droplets have been appeared). One has simply
require that $\Phi(z=0)$ isn't too far from the
maximum of supersaturation. This fact can be
also analytically proved here.

When
$x_{lim} = \infty$ as
 one can get the half-width of spectrum
 by condition
$$
\Delta z = 1
$$

\subsection{Pulse regime}

When $x_{lim} \gg 1$ the process of nucleation can
be described rather simple. To describe the first
peak
of the nucleation intensity
we have to solve equation without limit of
growth. This case is well known and can be solved
both by iterations \cite{Kuni} and by universal
solution \cite{PhysRevE94}. Thus, we get $N_1$ -
the number of droplets formed in the first peak.
One can see $N_1 \sim 1$.
 Then the
supersaturation $\zeta$ continues to fall until $z
\sim x_{lim}$.
The inequality $x_{lim} \gg 1$
allows here to speak
about the well formed peak
of nucleation intensity
and to use the standard
procedure of unlimited growth.
At $z \sim x_{lim}$ the
supesaturation begins to grow. At $z \sim
x_{lim}$ we have
$ f  \sim \exp(x_{lim}
- N_1 x_{lim}^3) \ll 1 $.
At $z \sim N_1 x_{lim}^3$ the second peak of
nucleation begins.

The second peak of nucleation can be described
absolutely analogously to the first one. So, we
have the set of identical peaks. The nucleation
occurs in the pulse regime.

This regime is shown in Figure 3.

\subsection{Smooth regime}

The opposite situation when $z_{lim} \ll 1$ can
be also solved analytically. Here kinetic equation
can be written as
\begin{equation}\label{k3}
\Phi(z) = \zeta (z) +
A \nu_{lim} \int_{-\infty}^{z-z_{lim}}
\exp(p(\zeta (x)- \Phi_* )) dx
\end{equation}
and
\begin{equation}\label{k31}
\Phi(z) = \zeta (z) +
A \nu_{lim} \int_{-\infty}^{z}
\exp(p(\zeta (x)- \Phi_* )) dx
\end{equation}

After differentiation with account of
linearization
we have
$$
\frac{dg}{dz} =
  A \nu_{lim}\exp( b z - p g(z))
$$
In our units $p=1$, $ A \nu_{lim} = 1$, $b=1$.
Integration with initial conditions
$$
g(z=-\infty) = 0
$$
gives
$$
\exp(g(z)) = A\nu_{lim}\exp(z)  +1
$$
This value will be marked as $g_{st}$.
Then the spectrum $f$ will be
\begin{equation}\label{stat}
f =  f_{st} \equiv \exp(x - \ln(\exp(x) +1))
\end{equation}
At
$x \rightarrow  \infty$
we see
$f_{st} \rightarrow f_{st\ lim} \equiv 1$.

This spectrum $f_{st}$ and the example
 of  numerical solution
% at
%$z_{lim} = 0.2$
are drawn in Figure 4.

These limit situations will form the base for
description of nucleation in all situations. But
here we have to stress two features:
\begin{itemize}
\item
The limit solutions have to be radically
modified. The force of the already presented
methods for these
solutions isn't sufficient to give the total
description.
\item
Description of the base of the iteration method can
not be suitable because the process isn't
restricted in time.
\end{itemize}

At first we shall present the general structure
of methods and then we shall show how to
determine the elements included in these
approaches.

\subsection{Advanced pulse regime}

To describe situations with
$x_{lim} \geq 2$ and $|x_{lim}
-
2|
\ll 2$
 one
has to use
the rescaled peaks  approach. This
approach is based on the following simple
considerations.

Due to the previous analysis we shall approximate
the spectrum by several peaks. Every peak is
formed by some external effective source
(initiated both by external conditions and by the
action of the tail of the previous peak).
The distance between peaks of nucleation
is
approximately
$$
\Delta_f x = N x_{lim}^3
$$
where $N$ is the number of droplets formed in the
previous peak.
Certainly, between different peaks the distance
can be different because $N$ will be different,
but the suitable approximation is to consider $N$
to be the ideal value, i.e. at the first peak.
The last formula
will be  rather evident if we notice that the
action of external conditions has simply to
compensate the loss  of substance which is equal
to $N x_{lim}^3$.

Here
we shall present a recurrent procedure
to calculate peaks
of nucleation. Suppose that we have
described the current
peak and know the coordinate of maximum $x_m$.

Now we have to rescale the amplitude of the next
peak. We neglect the non-linear behavior of
effective source. Then the next peak will be
similar after rescaling to the previous one. But
without rescaling in the same units the next peak
will have a new (ordinary smaller) height and
another width (ordinary wider).
So, now we have
to determine the new intensity $b'$ of the
effective source (the initial intensity was
$b_0=1$).

We know the  expression for intensity of vapor
consumption by droplets i.e. $dg / dz$:
$$
\frac{dg}{dz} =
3 A \int_{z-x_{lim}}^z (z-x)^2 \exp(x-g(x)) dx
$$
The subintegral function has the form
$$(z-x)^2 \exp(x-g(x))$$
and approximately has   maximum at
$(z-x) = \rho_a = 2$ for $x_{lim} >2$ and
$(z-x) = \rho_a = x_{lim}$
for $x_{lim} <2$ (here we can put
$z=0$ at maximum because this choice will result
only in parameters but not in the spectrum form).

The current peak have
coordinate $x_m$, the next peak will
have coordinate $x_m + \Delta_f x$. The maximum of
subintegral expression
for $ d \rho / d z $ for the next
peak will be attained
at $x_b = x_m +\Delta_f x - x_a$.The
main effect is connected
with a property that at $x=x_b$
the essential quantity  of
droplets from the previous peak
has attained the value $x_{lim}$.
But not all droplets have attained this value.
So, the
supersaturation $\zeta$ lies higher
than the supposed value
$$
b_0 z - N x_{lim}^3
$$
which would  ensure the
similarity of the next peak to the
previous one.

The supersaturation at
the corresponding  coordinate
$x_c = x_m - x_a$ for the
previous peak  was lower than at
the current peak.

We suppose that at
$
x= x_b +x_a = x_m + \Delta_f x
$
approximately corresponding  to the
coordinate of the next
peak all droplets of the previous
peak will attain the limit
value
$N x_{lim}^3$ and there will be  no difference
between the
value of effective ideal supersaturation  (the
supersaturation
imaginary formed without droplets of the
last peak taken into account) $\Omega$
at $x_m$ and at $x_m+\Delta_f x $
$$
\Omega (x_m) =
\Omega (x_m +\Delta_f x)
$$
The effective
intensity $b'$ of  the external source can be
approximated as
$$
b' \sim
\frac{\Omega (x_m + \Delta_f x ) -
\Omega (x_m +\Delta_f x
- x_a)}
{x_a}
$$
and have to be compared with
$$
b_0 \sim
\frac{\Omega (x_m  ) - \Omega (x_m
- x_a)}
{x_a}
$$

Then taking into account
$$
|\Omega ( x_m + \Delta_f x - x_m ) -
\Omega (x_m + \Delta_f x ) |
\ll
1
$$
and
$$
| \Omega (x_m) -
\Omega (x_m - \Omega (x_m - x_a) | \sim
x_a \geq 1
$$
one can come to
$$
b' = b_0 - \frac{\Delta \zeta }{x_a}
$$
where
$$
\Delta \zeta =
\Omega (x_m + \Delta_f x - x_a)
-
\Omega (x_m - x_a)
$$

Since $b_0 = 1$
one can come to
$$
b' = 1- \frac{\Delta \zeta }{x_a}
$$

We can express $\Delta \zeta $ as
$$
\Delta \zeta =
\ln \tilde{f} (x_b) - \ln \tilde{f}(x_c) =
\ln \frac{\tilde{f}(x_b)}{\tilde{f}(x_c)}
$$
where $\tilde{f}$ is the spectrum
corresponding to the
external ideal supersaturation.
Since $x_a > 1$ one can see
that  the effect of the influence
of the current peak is
negligible and
$$
\tilde{f} (x_b) = f(x_b)
$$
$$
\tilde{f} (x_c) = f(x_c)
$$

So, finally we have the following
formula for $b'$:
$$
b' = b_0 (1-\frac{\zeta_a}{x_a}) b
$$
where
$\zeta_a$ is the difference of supersaturations
at $x=x_c \equiv x_m-x_a$
and
at $x=x_b \equiv x_m-x_a+\Delta_f  x $ calculated
for the previous peak.
Here $x_m$ is the coordinate of the maximum of the
supersaturation.
The difference  $x_b - x_c$ can be calculated as
$$
x_b - x_c = \ln(\frac{f(x_b)}{f(x_c)})
$$

This completes the procedure.
The first peak
calculations have to be performed in explicit
manner.

The calculations are presented in Figure 5. We
see that the calculated peaks are very close to the
real solution.

Even such simple rescaling brings a rather
accurate result. When $x_{lim}>2$ the accuracy
will be better.
It can be proved analytically.
The simple account of variations
of $\Delta_f x $
will give the better accuracy also.
All other errors can be taken into account by the
standard perturbation technique.

\subsection{Advanced smooth regime}

We see that already at $x_{lim} = 2$ the first
peak lies not so far from the stationary solution
(\ref{stat}). Later all peaks will be even smaller
than the first one. So, we can consider
$(f-f_{st})/f_{st}$ as the small parameter and
linearize the kinetic equation over this
parameter.
Then this solution can be solved analytically and
will give the  oscillations relaxing to $f_{st}$.
We can avoid these long analytical derivations and
simply draw the relaxation oscillations
$$
f_{osc} = f_{st} + k_1 \exp((x-x_m)/k_2)
\cos((x-x_m)/\Delta_f x )
$$
with two parameters $k_1$ and $k_2$. The
arguments for derivation of $\Delta_f x $ takes
place here also.

Parameters $k_1$ and $k_2$ can be determined by a
coincidence of approximate and real solution at
maxima of two first peaks. We
choose maxima for coincidence of
real solution and  approximate solution
 because namely here the
intensity of nucleation has maxima.
The first local minima demonstrates the deviation
of approximate solution from the real one but
here the intensity of nucleation is small.

This solution is drawn in Figure 6. One can see
that precise solution is very close to
approximate one.

\subsection{Iterations to determine parameters}

Now we know the methods to describe nucleation
adequately, but parameters in these approximations
are unknown and our next task will be to determine
these parameters. To do this we have to know
the
evolution during several first peaks of
nucleation. In the advanced pulse method we have
to know the evolution during the first peak
including  the back side of the peak
and in
the advanced relaxation method we have to know the
positions and values of two first maximums. These
characteristics will be determined on the base of
iteration methods. But here the iteration methods
can not be got by direct generalization of already
presented procedures in the case of condensation
under the pure free molecular regime of vapor
consumption \cite{Kuni}, \cite{PhysRevE94}. Really,
even in the pure free molecular regime it is
difficult to describe even the back side of
spectrum on the base of standard iterations and
here we need to describe the second peak. The
limitation of growth will also diminish the
converging power of iterations. So we need to
reexamine the iteration procedure.

The evolution equation can be rewritten as
 $$
g =
$$
$$
\Theta(z - z_{lim})\{
 \int_{z-z_{lim}}^z (z-x)^3  \exp(x - g(x)) dx +
 \nu_{lim} \int_0^{z-z_{lim}}   \exp(x- g(x)) dx
\}+
$$
$$
\Theta(z_{lim}-z)
 \int_0^z (z-x)^3  \exp(x - g(x)) dx
$$

The initial
standard
iteration procedure can be determined
as
$$
g_{i+1} = \Theta(z - z_{lim})(
 \int_{z-z_{lim}}^z (z-x)^3  \exp(x - g_i(x)) dx +
 \nu_{lim} \int_0^{z-z_{lim}}   \exp(x- g_i(x)) dx
)+
$$
$$
\Theta(z_{lim}-z)
 \int_0^z (z-x)^3  \exp(x - g_i(x)) dx
$$

It will converge at every initial approximation.

As suitable initial approximations one can propose
$$
g_0 = 0
$$
or
$$
g_0 = g_{st}
$$
$$
g_{st} = 0 \ \ at \ \ x< \ln(f_{st\ lim }),
\ \ g_{st} = x \ \ at \ \ x \geq \ln(f_{st\ lim}) .
$$
Since here it will be impossible to take only two
first iterations there is no special significance
what approximation we shall use. We prefer to use
the first one because here the chain of
inequalities
\begin{equation} \label{chain}
g_0 < g_2 < ... < g_{2i} < ... <
g < ... < g_{2i+1} < ... < g_3< g_1
\end{equation}
for every fixed $x$ is observed.
These iterations are shown in Figure 7.

{\bf Decomposition in iterations }

The analytical calculation of iterations will stop
at the second step. For the unlimited free
molecular regime this number of iteration is
sufficient, but here we have to
 calculate the next iterations.
 It is possible analytically if we choose
 the  decomposition
of exponent in the subintegral function. But here the
following difficulty appear. The value of $g$ goes
to infinity and it isn't possible to decompose
$\exp(-g(x))$ over $g(x)$. The other possibility
is to decompose $\exp(x-g(x))$ over $x-g$ near
some characteristic value of the exponent
argument. But at $x \rightarrow -\infty$ the value
of $g$ goes to zero and the spectrum looks like
$\exp(-|x|)$. Then it is impossible to use the
second decomposition at initial moments of time
(i.e. at negative $x$). But this very period
drives the evolution during the first peak of
nucleation and, thus, plays the main role in
nucleation description. So, the pure second
approach can not lead to a suitable result.
So, we
need to give here more sophisticated approach.

We shall act in the following manner. Already the
first iteration gives us the approximate position
of the first maximum $x_{max\ 1}$ and the value of
this maximum $f_{max\ 1}$. Then we propose the
following procedure
\begin{itemize}
\item
When $x<x_{max\ 1}$ then
$$
\exp(-g) \approx \sum_{i=0}^4 \frac{(-g)^i}{i!}
$$
Here the first four
terms are taken into account. It is
sufficient to ensure the high relative accuracy.
\item
When $x \geq x_{max\ 1}$ then
$$
\exp(x-g)   \approx
\exp(\zeta_{base}) \sum_{i=0}^4 \frac{(x-g-
\zeta_{base})^i}{i!}
$$
where
$$
\zeta_{base} = \ln(\frac{f_{max\ 1} +
f_{st\ lim}}{2})
$$
is the base of decompositions.

Here decomposition of the corresponding exponent
 is limited
also by the first four
terms.

\end{itemize}

Then the iterations can be calculated
analytically at every step and
they give the adequate
approximation for the solution.
They converge to the solution of
kinetic equation where instead of $\exp$ one
should write the presented approximation.
Already the four first terms of decomposition
of the corresponding exponent
ensure the high accuracy.

These iterations are shown in Figure 8.

We see that already the several first iterations
give the practically precise positions of the
first maximum and the first minimum $f_{min\ 1}$
at $x_{min\ 1}$. Later
we see that iterations converge but every new
iteration gives the "step" of coincidence with the
real solution which becomes smaller and smaller.
The reason of this failure is the following:
really, the precise solution doesn't go far from
$f_{st\ lim}$ and decomposition works, but every
iteration inevitably goes to zero or to infinity
and on every iteration the decomposition does't
work.
This error leads to crisis
in convergence of iterations. So, we need to
modify iterations.

{\bf Cut-off of iterations }

Now we know $f_{max\ 1}$ and $f_{min\ 1}$ and can
use these characteristics. One can analytically
prove that for every $x>x_{max\ 1}$
$$
f_{min\ 1}  < f < f_{max\ 1}
$$
Then we can require that before we use
decomposition the following cut-off
\begin{itemize}
\item
When $\exp(x-g_i(x))>f_{max\ 1}$ we should take
instead of $\exp(x-g_i(x))$ the value the constant
$f_{max\ 1}$ and fulfill no further
decompositions\footnote{One can see
that  iterations  escape
from the real solution.}.

\item
When $\exp(x-g_i(x))< f_{min\ 1}$ we should take
instead of $\exp(x-g_i(x))$ the value the constant
$f_{min\ 1}$ and fulfill no further
decompositions.

\item
When $f_{min\ 1} < \exp(x-g_i(x))< f_{max\ 1}$ no
special actions are required and we have to make
the mentioned decompositions.

\end{itemize}

The convergence of new iterations is higher than
the convergence of previous iterations.
New iterations
evidently converge to the real
solution\footnote{with corresponding decomposition.}.
Also it is very easy to
estimate rigorously the error of decomposition
because now we have an estimate $f_{min \ 1} \leq
f_i \leq f_{max\ 1}$ for every $x$ and for every
$i$.

To complete the description of iterations one can
prove the following important statements

\begin{itemize}

\item
The cut-off iterations without decompositions
converge to the real solution of kinetic equation
and they converge faster than the first-type
iterations. The chains of inequalities
(\ref{chain}) remain valid for the cut-off
iterations without decompositions.

\item
The iterations with decompositions converge to the
solution of kinetic equation where function $\exp$
is treated according to the mentioned
decompositions. The chains of inequalities
(\ref{chain}) remain valid (here $g$ is solution
of kinetic equation where the function $\exp$ is
treated according to the mentioned
decompositions)\footnote{To show this we have to
see that decomposition of $\exp$ conserves the
monotonious properties. It can be shown if we
differentiate the approximation and note that the
derivative of approximation is the approximation
for the derivative of exponent.
 Then with the necessary terms
taken into account  and with
account that $f_{min \ 1} \leq f_i \leq f_{max\
1}$ this derivative can be made positive. Then the
monotonious properties
remain even after decompositions.}.

\end{itemize}

For practical needs one can show that for $x_{lim}
\leq 2$ already the  sixth iteration
ensures
the correct values of $f_{max\ 1}, x_{max\ 2}$ and
$f_{max_2}, x_{max\ 2}$ (the value and
the coordinate of the second peak)\footnote{When
 $x_{lim}<2$
then the
iterations will approximate these values even
better then at $x_{lim} = 2$.
It can be proved analytically. }. This is
all we need to construct the relaxation
oscillations.

Behavior of the cut-off iterations
is shown in Figure 9.
One can see the high rate of convergence.

When $x_{lim}>2$ we can use the advanced pulse
method and here we need to describe only the
first peak. It can be done by the first four
iterations (Really, if it would be necessary to
get the coordinate of the second peak we might
have difficulties because of the very small
value of $f_{min\ 1})$.

This solves the problem to get
the parameters of approximation and
completes the  description of
the nucleation under
  the break of the blow-up growth.

\pagebreak

\input break1.txt

\pagebreak

\begin{picture}(300,300)
\put(25,25){\line(0,1){250}}
\put(25,25){\line(1,0){250}}
\put(275,275){\line(-1,0){250}}
\put(275,275){\line(0,-1){250}}
\put(50,50){\vector(1,0){200}}
\put(50,50){\vector(0,1){200}}
\put(40,40){$0$}
\put(125,40){$1$}
\put(125,50){\line(0,1){1}}
\put(210,40){$x_{lim}$}
\put(30,200){$0.05$}
\put(50,200){\line(0,1){1}}
\put(60,230){$\epsilon$}
\put(51.500000,50.000572){\special{em:moveto}}
\put(53.000000,50.005444){\special{em:lineto}}
\put(54.500000,50.018391){\special{em:lineto}}
\put(56.000000,50.042480){\special{em:lineto}}
\put(57.500000,50.073215){\special{em:lineto}}
\put(59.000000,50.106453){\special{em:lineto}}
\put(60.500000,50.116566){\special{em:lineto}}
\put(62.000000,50.069698){\special{em:lineto}}
\put(63.500000,50.100639){\special{em:lineto}}
\put(65.000000,50.439247){\special{em:lineto}}
\put(66.500000,51.046631){\special{em:lineto}}
\put(68.000000,52.002064){\special{em:lineto}}
\put(69.500000,53.377995){\special{em:lineto}}
\put(71.000000,55.236561){\special{em:lineto}}
\put(72.500000,57.632820){\special{em:lineto}}
\put(74.000000,60.591881){\special{em:lineto}}
\put(75.500000,64.109848){\special{em:lineto}}
\put(77.000000,68.184868){\special{em:lineto}}
\put(78.500008,72.781731){\special{em:lineto}}
\put(80.000008,77.868019){\special{em:lineto}}
\put(81.500008,83.369209){\special{em:lineto}}
\put(83.000008,89.224373){\special{em:lineto}}
\put(84.500008,95.363220){\special{em:lineto}}
\put(86.000008,101.692696){\special{em:lineto}}
\put(87.500008,108.131332){\special{em:lineto}}
\put(89.000008,114.588226){\special{em:lineto}}
\put(90.500008,120.970375){\special{em:lineto}}
\put(92.000008,127.182907){\special{em:lineto}}
\put(93.500000,133.133469){\special{em:lineto}}
\put(95.000000,138.736542){\special{em:lineto}}
\put(96.500000,143.908890){\special{em:lineto}}
\put(98.000000,148.436249){\special{em:lineto}}
\put(99.500000,152.477142){\special{em:lineto}}
\put(100.999992,155.880096){\special{em:lineto}}
\put(102.499992,158.599304){\special{em:lineto}}
\put(103.999992,160.596863){\special{em:lineto}}
\put(105.499992,161.870117){\special{em:lineto}}
\put(106.999992,162.402969){\special{em:lineto}}
\put(108.499992,162.215500){\special{em:lineto}}
\put(109.999985,161.365387){\special{em:lineto}}
\put(111.499985,159.861252){\special{em:lineto}}
\put(112.999985,157.780640){\special{em:lineto}}
\put(114.499985,155.162094){\special{em:lineto}}
\put(115.999985,152.071793){\special{em:lineto}}
\put(117.499977,148.570740){\special{em:lineto}}
\put(118.999977,144.666153){\special{em:lineto}}
\put(120.499977,140.359573){\special{em:lineto}}
\put(121.999977,135.670258){\special{em:lineto}}
\put(123.499977,130.532013){\special{em:lineto}}
\put(124.999969,124.910736){\special{em:lineto}}
\put(126.499969,118.755692){\special{em:lineto}}
\put(127.999969,112.035133){\special{em:lineto}}
\put(129.499969,104.749809){\special{em:lineto}}
\put(130.999969,96.921379){\special{em:lineto}}
\put(132.499969,88.663719){\special{em:lineto}}
\put(133.999969,80.123917){\special{em:lineto}}
\put(135.499969,71.482384){\special{em:lineto}}
\put(136.999954,62.956280){\special{em:lineto}}
\put(138.499954,54.776180){\special{em:lineto}}
\put(139.999954,52.868370){\special{em:lineto}}
\put(141.499954,59.820068){\special{em:lineto}}
\put(142.999954,65.974335){\special{em:lineto}}
\put(144.499954,71.232216){\special{em:lineto}}
\put(145.999954,75.615059){\special{em:lineto}}
\put(147.499954,79.112961){\special{em:lineto}}
\put(148.999954,81.859444){\special{em:lineto}}
\put(150.499954,83.869644){\special{em:lineto}}
\put(151.999954,85.355637){\special{em:lineto}}
\put(153.499939,86.571884){\special{em:lineto}}
\put(154.999939,87.152527){\special{em:lineto}}
\put(156.499939,87.265343){\special{em:lineto}}
\put(157.999939,87.423080){\special{em:lineto}}
\put(159.499939,87.461845){\special{em:lineto}}
\put(160.999939,87.475044){\special{em:lineto}}
\put(162.499939,87.118111){\special{em:lineto}}
\put(163.999939,86.970879){\special{em:lineto}}
\put(165.499939,86.806664){\special{em:lineto}}
\put(166.999939,86.667419){\special{em:lineto}}
\put(168.499939,86.688629){\special{em:lineto}}
\put(169.999924,86.552361){\special{em:lineto}}
\put(171.499924,86.453728){\special{em:lineto}}
\put(172.999924,86.384911){\special{em:lineto}}
\put(174.499924,86.336365){\special{em:lineto}}
\put(175.999924,86.303032){\special{em:lineto}}
\put(177.499924,86.280220){\special{em:lineto}}
\put(178.999924,86.265404){\special{em:lineto}}
\put(180.499924,86.254684){\special{em:lineto}}
\put(181.999924,86.247665){\special{em:lineto}}
\put(183.499924,86.243378){\special{em:lineto}}
\put(184.999908,86.240456){\special{em:lineto}}
\put(186.499908,86.238892){\special{em:lineto}}
\put(187.999908,86.237724){\special{em:lineto}}
\put(189.499908,86.236946){\special{em:lineto}}
\put(190.999908,86.236557){\special{em:lineto}}
\put(192.499908,86.236557){\special{em:lineto}}
\put(193.999908,86.236557){\special{em:lineto}}
\put(195.499908,86.236557){\special{em:lineto}}
\put(196.999908,86.236557){\special{em:lineto}}
\put(198.499908,86.236557){\special{em:lineto}}
\put(199.999908,86.236557){\special{em:lineto}}
\put(201.499908,86.236557){\special{em:lineto}}
\end{picture}

 Figure 2

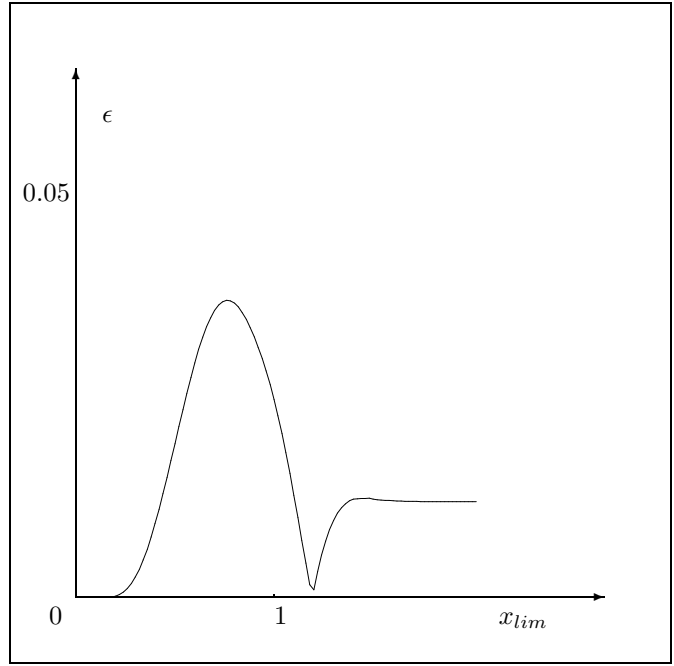
 Relative error $\epsilon$ of the special approximation

\pagebreak

\input bredyn3.txt

\pagebreak

\input bredyn4.txt

\pagebreak

\input bredyn5.txt

\pagebreak

\input break6.txt

\pagebreak

\input break7.txt

\pagebreak

\input break8.txt

\pagebreak

\input break9.txt

\pagebreak

 \end{document}